# On summation of 3*j* - Wigner symbols

Alexey N. Hopersky, Alexey M. Nadolinsky *and* Rustam V. Koneev

*Rostov State Transport University, 344038, Rostov-on-Don, Russia*

*E-mail*: qedhop@mail.ru, amnrnd@mail.ru, koneev@gmail.com

**Abstract**. Within the framework of the theory of irreducible tensor operators, using well-known general analytical results for double sums ($\Sigma_{jm}$) of products of two 3*j* - Wigner symbols, analytical expressions for single sums ($\Sigma_m$) for the values $j_1 = j_2 = 1$ and $j = 2$ parameters of the upper row 3*j* - Wigner symbol are specified. The expressions obtained supplement the well-known analytical results of the theory of angular momentum and are in demand in solving, in particular, such problems of atomic physics as the construction of a nonrelativistic quantum theory of single and double bremsstrahlung when a photon is scattered by an atom (atomic ion) (Hopersky *et al* [6,7,8]) and two-photon resonance single ionization of the deep shell of an atomic ion (Hopersky *et al* [9]).

**Introduction**

The theory of irreducible tensor operators is an integral part of modern mathematical analysis [1, 2]. When implementing the methods of the theory of irreducible tensor operators, for example, within the framework of quantum physics problems [3], mathematical structures arise for which, as far as we know, there are no specified *analytical* results in the published literature – summation formulas for the 3*j*-Wigner symbols [4] with a fixed total angular moment of the wave function of the state of a quantum system. Summation formulas arise when constructing the products of partial amplitudes of the probability of transitions between the states of the quantum system under study. The analytical structures of each of the probability amplitudes (matrix elements of the radiation and contact transition operators in the secondary quantization representation as multiple integrals of spatial and angular variables) are determined by the Wigner–Eckart theorem [4, 5]:

$$\left\langle JM \left| T_q^{(k)} \right| J'M' \right\rangle = (-1)^{J-M} \cdot \left( J \left\| T^{(k)} \right\| J' \right) \cdot \begin{pmatrix} J & k & J' \\ -M & q & M' \end{pmatrix}. \qquad (1)$$

In (1) the following are determined: $T_q^{(k)}$ – irreducible tensor operator of the transition between $|JM\rangle$ – and $|J'M'\rangle$ – states of a quantum system, $k$ – rank of the transition operator and its projection $q = -k, -k+1, ..., k$, $J(M)$ – and $J'(M')$ – total moments (their projections $M = -J, -J+1, ..., J$) of states and $\left( J \left\| T^{(k)} \right\| J' \right)$ – reduced (independent of projections) matrix element. According to (1), the dependence of the probability amplitude of the transition on the projections $M$, $q$ is $M'$ determined by the phase multiplier and the 3*j* –Wigner symbol. In this case, the 3*j*-Wigner symbols character is non-zero when the requirements for its strings are met: $J' = J + k$, $J + k - 1$, …, $|J - k|$ (triangle condition) and $q + M' = M$. In this article, we specify the analytic expressions for two summation formulas,

$$\sum_m (-1)^m \begin{pmatrix} j_1 & j_2 & j \\ a & b & m \end{pmatrix} \begin{pmatrix} j_1 & j_2 & j \\ c & d & -m \end{pmatrix}, \qquad (2)$$

$$\sum_m \begin{pmatrix} j_1 & j_2 & j \\ a & b & m \end{pmatrix} \begin{pmatrix} j_1 & j_2 & j \\ c & d & m \end{pmatrix}, \qquad (3)$$

in the most common case $j_1 = j_2 = 1$, $j = 2$. In sums (2), (3) and further, we unified the notations for the parameters 3*j* – Wigner symbol: $J \to j_1$, $k \to j_2$, $J' \to j$, $-M \to a$, $q \to b$ and $M' \to m$. So, for example, the accounting of dipole ($k = 1$, $q = -1, 0, 1$) transitions involving electrons of the continuous spectrum $p(J=1)$ – and $d(J=2)$ – symmetry of intermediate (virtual) and final (observable) states of inelastic scattering of a photon by a quantum system is accompanied by the following. Sum (3) arises when calculating the probability amplitude of the initiated one-time (transition between terms $^1P_1 \to {}^1D_2$) of bremsstrahlung radiation [6, 7], while the sum of (2) arises when calculating the probability amplitude of the initiated nonlocal double (transition between terms $^1P_1 \to {}^1D_2 \to {}^1P_1$) bremsstrahlung radiation [8] with inelastic scattering of a photon by an atom (atomic ion). Sum (3) also occurs in the construction of a generalized cross-section of two-photon resonance single ionization of the K-shell of an atomic ion [9].

**Results**

**Statement 1.** At $j_1 = j_2 = 1$ and $j = 2$ the sum (2) takes the form:

$$\sum_{m=-2}^{2} (-1)^m \cdot \begin{pmatrix} 1 & 1 & 2 \\ a & b & m \end{pmatrix} \begin{pmatrix} 1 & 1 & 2 \\ c & d & -m \end{pmatrix} =$$

$$= \frac{1}{5}(-1)^{d-a} \cdot \left[ \delta_{b,-d} \cdot \delta_{c,-a} - \frac{1}{2}\left(a \cdot d + \frac{2}{3}\right)\delta_{b,-a} \cdot \delta_{c,-d} - A \right], \tag{4}$$

$$A = \frac{1}{4}(A_+ + A_-),$$

$$A_+ = \sqrt{(2+a)(1-a)(1+d)(2-d)} \cdot \delta_{b,-a+1} \cdot \delta_{c,-d+1},$$

$$A_- = \sqrt{(2-a)(1+a)(1-d)(2+d)} \cdot \delta_{b,-a-1} \cdot \delta_{c,-d-1}.$$

**Proof.** Let's consider a well-known general analytical result (the orthogonality condition of $3j$ – Wigner symbols) [2]:

$$\sum_j \sum_m (-1)^{j-m} (2j+1) \begin{pmatrix} j_1 & j_2 & j \\ a & b & m \end{pmatrix} \begin{pmatrix} j_1 & j_2 & j \\ c & d & -m \end{pmatrix} =$$

$$= (-1)^{j_1+j_2-a-b} \cdot \delta_{a,-c} \cdot \delta_{b,-d}, \tag{5}$$

where $m = -j, -j+1, ..., j$ and $\delta_{\alpha,\beta}$ is the symbol of Kronecker–Weierstrass. In our case, $j_1 = j_2 = 1$, each of the integer $a$, $b$, $c$, and $d$ takes the values -1, 0, +1. Then, due to the requirement $a+b+m=0$ for the bottom line $3j$-Wigner symbol, the resulting total moment is constrained by values $j = 0, 1, 2$ and the expression (5) takes the form:

$$\frac{1}{3}(-1)^{b+d} \cdot \delta_{a,-b} \cdot \delta_{c,-d} + 3 \cdot \sum_{m=-1}^{1} (-1)^{1-m} \cdot \begin{pmatrix} 1 & 1 & 1 \\ a & b & m \end{pmatrix} \begin{pmatrix} 1 & 1 & 1 \\ c & d & -m \end{pmatrix} +$$

$$+ 5 \cdot \sum_{m=-2}^{2} (-1)^m \cdot \begin{pmatrix} 1 & 1 & 2 \\ a & b & m \end{pmatrix} \begin{pmatrix} 1 & 1 & 2 \\ c & d & -m \end{pmatrix} = (-1)^{a+b} \cdot \delta_{a,-c} \cdot \delta_{b,-d}. \tag{6}$$

In (6) the first term on the left side is equal to the product of $3j$ – Wigner symbols with a zero column,

$$\begin{pmatrix} 1 & 1 & 0 \\ a & b & 0 \end{pmatrix} \begin{pmatrix} 1 & 1 & 0 \\ c & d & 0 \end{pmatrix}, \quad \begin{pmatrix} 1 & 1 & 0 \\ a & b & 0 \end{pmatrix} = \frac{1}{\sqrt{3}} \cdot (-1)^{1+b} \cdot \delta_{a,-b},$$

and the sum of interest arises (2). Let's take into account the known results for $3j$ – Wigner symbols in the first sum of (6) [10, 11]:

$$\begin{pmatrix} 1 & 1 & 1 \\ a & b & -1 \end{pmatrix} = \frac{1}{2\sqrt{3}} \cdot (-1)^{1+b} \cdot \sqrt{(1+b)(2-b)} \cdot \delta_{a,1-b}, \tag{7}$$

$$\begin{pmatrix} 1 & 1 & 1 \\ a & b & 0 \end{pmatrix} = \frac{1}{\sqrt{6}} \cdot (-1)^{1+b} \cdot b \cdot \delta_{a,-b}, \tag{8}$$

$$\begin{pmatrix} 1 & 1 & 1 \\ a & b & 1 \end{pmatrix} = \frac{1}{2\sqrt{3}} \cdot (-1)^{b} \cdot \sqrt{(1-b)(2+b)} \cdot \delta_{a,-1-b}, \tag{9}$$

where the Kronecker–Weierstrass symbols reproduce the requirement $a+b+m=0$. Taking into account (7), (8) and (9) from (6) we get the sum (4). <u>Statement 1 has been proven.</u>

**Statement 2.** At $j_1 = j_2 = 1$ and $j = 2$ the sum (3) takes the form:

$$\sum_{m=-2}^{2} \begin{pmatrix} 1 & 1 & 2 \\ a & b & m \end{pmatrix} \begin{pmatrix} 1 & 1 & 2 \\ c & d & m \end{pmatrix} =$$
$$= \frac{1}{5} \cdot \delta_{a,c} \cdot \delta_{b,d} - \frac{1}{20} (-1)^{b+d} \cdot \left[ \left( 2bd + \frac{4}{3} \right) \cdot \delta_{a,-b} \cdot \delta_{c,-d} + B \right], \tag{10}$$
$$B = B_+ + B_-,$$
$$B_+ = \sqrt{(1+b)(2-b)(1+d)(2-d)} \cdot \delta_{a,1-b} \cdot \delta_{c,1-d},$$
$$B_- = \sqrt{(1-b)(2+b)(1-d)(2+d)} \cdot \delta_{a,-1-b} \cdot \delta_{c,-1-d}.$$

**Proof.** Let's consider a well-known general analytical result (the orthogonality condition of $3j$ – Wigner symbols) [2]:

$$\sum_j \sum_m (2j+1) \begin{pmatrix} j_1 & j_2 & j \\ a & b & m \end{pmatrix} \begin{pmatrix} j_1 & j_2 & j \\ c & d & m \end{pmatrix} = \delta_{a,c} \cdot \delta_{b,d}. \tag{11}$$

For $j_1 = j_2 = 1$ and $j = 0, 1, 2$ expression (11) takes the form:

$$\frac{1}{3} (-1)^{b+d} \cdot \delta_{a,-b} \cdot \delta_{c,-d} + 3 \cdot \sum_{m=-1}^{1} (-1)^{1-m} \cdot \begin{pmatrix} 1 & 1 & 1 \\ a & b & m \end{pmatrix} \begin{pmatrix} 1 & 1 & 1 \\ c & d & m \end{pmatrix} +$$
$$+ 5 \cdot \sum_{m=-2}^{2} \begin{pmatrix} 1 & 1 & 2 \\ a & b & m \end{pmatrix} \begin{pmatrix} 1 & 1 & 2 \\ c & d & m \end{pmatrix} = \delta_{a,c} \cdot \delta_{b,d}, \tag{12}$$

where the amount of interest arises (3). Taking into account (7), (8) and (9) from (12) we get the sum (10). <u>Statement 2 has been proven.</u>

**Comments**

**1.** The realization of the terms $A_+$, $A_-$, $B_+$ and $B_-$ in sums (4) and (10) when constructing cross-sections of the processes under study requires direct reference to spherical functions $Y_{kq}(\boldsymbol{e}_i)$ ($\boldsymbol{e}_i$ – unit polarization vector of $i$-photon in spherical coordinates, $k = 1$, $q = -1, 0, 1$) [2]. This, in turn, is accompanied by the need to initially fix the geometry (specify the analytical form of spherical functions) of the *proposed* physical experiment and to carry out rather cumbersome calculations [6,7,8,9]. However, it is unacceptable to keep only the first terms of sums (4) and (10) in the right-hand parts as an "approximation". In fact, for example, the "approximation" of the transition from the sum (10) to the expression

$$\sum_{m=-2}^{2} \begin{pmatrix} 1 & 1 & 2 \\ a & b & m \end{pmatrix} \begin{pmatrix} 1 & 1 & 2 \\ c & d & m \end{pmatrix} \simeq \frac{1}{5} \cdot \delta_{a,c} \cdot \delta_{b,d}$$

results in taking into account only the value of the projection of the total moment $m = 0$, while information about the contribution of the values $m = -2, -1, +1, +2$ is lost. This effect, in turn, from a physical point of view, leads to the loss of fundamental information about the polarization properties (for example, dependence on the scattering angle as the angle between the wave vectors of the incident and scattered photons) observed in the proposed experiment of the differential cross-section of the process under study.

**2.** Of course, the methods of **Statements 1, 2** can construct similar sums of products of 3$j$-Wigner symbols for supreme total moments $j \geq 3$ (see, for example, dipole transition between terms $^1D_2 \to {}^1F_3$). Such constructions are accompanied by a sharp increase in the volume of calculations and are the subject of future research.

*In this article, we present the English and extended version of the work* [12].